\documentclass[prd,%
preprint,%
tightenlines,%
showpacs,%
nofootinbib,%
superscriptaddress,%
amsfonts,%
]{revtex4}

\begin{document}

\preprint{UFIFT-QG-14-01}

\title{Field equations and cosmology for a class\\
of nonlocal metric models of MOND}

\author{C\'edric~Deffayet} \email{deffayet@iap.fr}
\affiliation{UPMC-CNRS, UMR7095, Institut d'Astrophysique de Paris,\\
${\mathcal{G}}{\mathbb{R}}\varepsilon{\mathbb{C}}{\mathcal{O}}$,
98bis boulevard Arago, F-75014 Paris, France}
\affiliation{IHES, Le Bois-Marie, 35 route de Chartres,
F-91440 Bures-sur-Yvette, France}

\author{Gilles~\surname{Esposito-Far\`ese}} \email{gef@iap.fr}
\affiliation{UPMC-CNRS, UMR7095, Institut d'Astrophysique de Paris,\\
${\mathcal{G}}{\mathbb{R}}\varepsilon{\mathbb{C}}{\mathcal{O}}$,
98bis boulevard Arago, F-75014 Paris, France}

\author{Richard P.~Woodard} \email{woodard@phys.ufl.edu}
\affiliation{Department of Physics, University of Florida,
FL-32611 Gainesville, USA}

\begin{abstract}
We consider a class of nonlocal, pure-metric modified gravity
models which were developed to reproduce the Tully-Fisher
relation without dark matter and without changing the amount of
weak lensing predicted by general relativity. Previous work gave
only the weak field limiting form of the field equations
specialized to a static and spherically symmetric geometry. Here
we derive the full field equations and specialize them to a
homogeneous, isotropic and spatially flat geometry. We also
discuss the problem of fitting the free function to reproduce the
expansion history. Results are derived for models in which the
MOND acceleration $a_0 \approx 1.2 \times
10^{-10}~\text{m}.\text{s}^{-2}$ is a fundamental constant and
for the more phenomenologically interesting case in which the
MOND acceleration changes with the cosmological expansion rate.
\end{abstract}

\pacs{04.50.Kd, 95.35.+d, 98.62.-g\\
\\
\textit{Dedicated to Stanley Deser on the occasion of his 83rd
birthday.}}

\maketitle

\section{Introduction}\label{intro}
Despite the successes of the standard model of cosmology based on
general relativity, many feel unsatisfied that the only currently
available evidences for dark matter and dark energy are indirect,
and it is certainly worth pursuing other approaches. One of them
is the MOND paradigm (standing for MOdified Newtonian Dynamics),
as proposed by Milgrom \cite{Milgrom1}, which lead to many
successful explanations of various observations as well as to
predictions which were confirmed \cite{Milgrom3}. In particular,
it explains the Tully-Fisher relation \cite{TF}, which states
that the observed limiting rotation velocity of galaxies,
$v_{\infty}$, scales as the fourth root of the baryonic mass of
the galaxy (see \cite{Stacy} for a recent confirmation of this
relation). As it was first formulated in a non relativistic way,
Milgrom's proposal stipulates that a test particle at a distance
$r$ from a mass $M$ will experience a gravitational acceleration
given by the Newtonian expression $a_N = GM/r^2$ as long as $a_N$
is (much) larger than a critical acceleration $a_0$, while the
same particle will undergo the MOND acceleration $a_{\text{MOND}}
= \sqrt{a_N a_0} = \sqrt{GM a_0}/r$ when $a_N$ is smaller than
$a_0$. A constant value for $a_0$ of about $1.2 \times 10^{-10}\,
\text{m}.\text{s}^{-2}$ leads to good fits of galaxy rotation
curves using reasonable mass-to-luminosity ratios \cite{BdJ},
without the need for non-baryonic dark matter \cite{SanMc} (see
however \cite{Randriamampandry:2014eoa}).
As noticed by many authors, this numerical value for $a_0$ is very
close to $c H_0/2\pi$, where $H_0$ is the current value of the
Hubble parameter. This gives some support to the idea that the
MOND parameter $a_0$ actually varies with time over the
cosmological history of the Universe \cite{dynaccel}, and we will
discuss some aspects of this possibility in this work (together
with the constant $a_0$ case). Such a time variation of $a_0$
could also lead to some specific observational signatures
\cite{dynaccelobs}.

Despite its successes, the current MOND framework also suffers
from various problems both at the level of observation fitting
and theoretical construction. It is indeed well known that some
amount of dark matter is needed (or possibly a deviation from the
original MOND formulation without dark matter) in order for MOND
to fit velocity dispersion in galaxy clusters. The bullet cluster
\cite{Clowe:2003tk} is also often presented as a serious puzzle
for MOND (see however
\cite{Dai:2008sf,Feix:2007zm,Angus:2007qj}).
On the theory side, the great challenge has been to construct a
relativistic extension of MOND that reproduces, without dark
matter, both the observed cosmology and the observed amount of
weak lensing. Some attempts along this line include TeVeS
theories
\cite{Bekenstein:1992pj,Bekenstein:1993fs,Sanders:1996wk,
TeVeS,Bekenstein:2004ca,Sanders:2005vd,Sagi:2009kd,Bekenstein:2010pt}
and other models with scalar and vector fields
\cite{Moffat:2005si,Zlosnik:2006zu}, Milgrom's bi-metric model
\cite{Milgrom:2009gv}, and nonlocal, metric-based models
\cite{Hehl:2008eu,Hehl:2009es,Blome:2010xn,Arraut:2013qra}. It is
fair to say that these attempts, however very interesting, need
to be further explored and consolidated.

In our previous work \cite{Deffayet:2011sk}, we introduced yet a
new relativistic formulation of MOND which is at the root of the
present work. In this formulation, the only dynamical degrees of
freedom are those of a metric. We called hence there such a
theory a ``pure-metric'' theory, in contrast to the TeVeS model or
bi-metric theory which both contain degrees of freedom which are
added explicitly to those of the metric. Note that such a
distinction cannot always be considered as very deep: e.g., it
is well known that $f(R)$ theories (which would qualify as a pure
metric theory using our terminology) can also be formulated as
scalar-tensor theories, i.e., as theories with a metric and an
extra scalar in the gravitational sector. We will in turn use
here sometimes scalars to describe our model. However our scalars
will not have a proper dynamics as we will explain later. The
advantages of a pure metric based theory are that it allows a
clear way to build the matter coupling in agreement with the
equivalence principle as well as a simple comparison with general
relativity. As argued in particular in
\cite{SW2,Deffayet:2011sk}, it can however be shown that a
pure-metric based theory of MOND has to be non local, and this is
the case of the theory hereunder consideration. The point of this
paper is to further develop a class of generally coordinate
invariant, nonlocal metric realizations of MOND which have been
proposed in \cite{Deffayet:2011sk}.

Nonlocal metric extensions of gravity (irrespective of MOND) have
been much studied \cite{nonloc} because they offer a richer
phenomenology than $f(R)$ models \cite{Mark,NO}, which are the
only local, invariant, metric-based and kinetically stable
extensions of general relativity \cite{RPW,Woodard:2014wia}.
We do not believe fundamental theory is nonlocal, but rather that
nonlocal extensions of general relativity derive from quantum
infrared corrections to the effective field equations that became
nonperturbatively strong during an extended phase of primordial
inflation \cite{TWa}. Put simply, we believe that MOND derives
from the gravitational vacuum polarization of the vast ensemble
of infrared gravitons created during primordial inflation.
Although our class of models is, at this stage, purely
phenomenological, our suspicion about its probable origin helps
to justify two features which would otherwise be inexplicable:
\begin{itemize}
\item{Our models possess an initial time $t_i$; and}
\item{Our models predict significant deviations from general
relativity on large scales but not on small scales.}
\end{itemize}

Our previous work \cite{Deffayet:2011sk} began by deriving
phenomenological equations which any metric-based theory of MOND
must obey for static, spherically symmetric geometries of the
form,
\begin{equation}
ds^2 \equiv g_{\mu\nu} dx^{\mu} dx^{\nu} = -\Bigl[1 + b(r)
\Bigr] dt^2 + \Bigl[1 + a(r)\Bigr] dr^2 + r^2 d\Omega^2 \; .
\end{equation}
If the energy density $\rho(r)$ is such that the system is
everywhere in the MOND regime (as would be the case, for example,
in a low surface brightness galaxy) then the Tully-Fisher
relation implies \cite{Deffayet:2011sk},
\begin{equation}
\frac1{2 a_0 r^2} \frac{d}{dr} \Bigl( [r b'(r)]^2 \Bigr)
= 8\pi G \rho(r) \; . \label{g00}
\end{equation}
We fixed the other potential $a(r)$ by requiring that weak
lensing agrees exactly with what general relativity predicts
assuming the potential $b(r)$ is known,
\begin{equation}
r b'(r) - a(r) = 0 \; . \label{grr}
\end{equation}
Note that any power of (\ref{grr}) would work as well because the
right hand side of the equation vanishes. If one
allows only an approximate agreement with general relativity as
far as lensing is concerned, as e.g. is allowed by cosmological
data using lensing, a coefficient of order one can just be
inserted in front of $a(r)$ in the equation above.

We assumed a Lagrangian consisting of general relativity and
normal matter, plus a MOND correction term $\Delta \mathcal{L}$,
\begin{equation}
\mathcal{L} = \frac{R \sqrt{-g}}{16 \pi G} + \Delta \mathcal{L}
+ \mathcal{L}_{\rm matter} \; .
\end{equation}
Because the matter Lagrangian is unchanged from general
relativity (except for the absence of dark matter) our field
equations take the form,
\begin{equation}
G_{\mu\nu} + \Delta G_{\mu\nu} = 8\pi G T_{\mu\nu} \; ,
\label{fieldeqn}
\end{equation}
where $G_{\mu\nu} \equiv R_{\mu\nu} - \frac12 g_{\mu\nu} R$ is
the usual Einstein tensor and $T_{\mu\nu}$ is the usual
stress-energy tensor. (We employ a metric with mostly plus
signature with Riemann tensor $R^{\rho}_{~\sigma\mu\nu} = +
\partial_{\mu} \Gamma^{\rho}_{~\nu\sigma} - \dots$ and Ricci
tensor $R_{\mu\nu} \equiv R^{\rho}_{~\mu\rho\nu}$.) The MOND
correction $\Delta G_{\mu\nu}$ to the Einstein tensor comes from
varying the action deriving from $\Delta \mathcal{L}$,
namely the spacetime integral $\Delta S$ of $\Delta
\mathcal{L}$. We get
\begin{equation}
\Delta G_{\mu\nu}(x) \equiv \frac{16 \pi G}{\sqrt{-g}}
\frac{\delta \Delta S[g]}{\delta g^{\mu\nu}(x)} \; . \label{DGmn}
\end{equation}
We constructed the MOND correction $\Delta \mathcal{L}$ such
that, in the static, spherically symmetric and ultra-weak field
regime, the $\mu = \nu = 0$ equation reduces to (\ref{g00}) and
the $\mu = \nu = r$ equation is proportional to (\ref{grr}).

We found that it sufficed to employ a single nonlocal scalar,
\begin{equation}
Y[g] \equiv g^{\mu\nu} \partial_{\mu} \frac2{\Box} \Bigl[
u^{\alpha} u^{\beta} R_{\alpha\beta} \Bigr] \partial_{\nu}
\frac2{\Box} \Bigl[ u^{\rho} u^{\sigma} R_{\rho \sigma}
\Bigr] \; . \label{Ydef}
\end{equation}
Here and henceforth $\Box \equiv \frac1{\sqrt{-g}} \partial_{\mu}
(\sqrt{-g} g^{\mu\nu} \partial_{\nu})$ is the covariant scalar
d'Alem\-bert\-ian, and its inverse is defined with retarded
boundary conditions on the initial value surface. The timelike
4-velocity field $u^{\mu}[g]$ is the normalized gradient of some
nonlocal scalar functional of the metric $\chi[g]$, such as the
invariant volume of the past lightcone, which grows in the
timelike direction,
\begin{equation}
u^{\mu}[g] \equiv \frac{-g^{\mu\nu} \partial_{\nu} \chi[g]}{
\sqrt{-g^{\alpha\beta} \partial_{\alpha} \chi[g]
\partial_{\beta} \chi[g]}} \; . \label{udef}
\end{equation}
In the static, spherically symmetric and ultra-weak field limit,
$Y[g]$ reduces to just $[b'(r)]^2$ and we could reproduce
(\ref{g00}), without disturbing the general relativistic relation
(\ref{grr}), with a MOND addition of the form,
\begin{equation}
\Delta \mathcal{L}_y = \frac{a_0^2}{16 \pi G} \Biggl\{ \frac12
\Bigl( \frac{Y}{a_0^2} \Bigr) - \frac16 \Bigl( \frac{Y}{a_0^2}
\Bigr)^{\frac32} + \dots \Biggr\} \sqrt{-g} \; . \label{DLY}
\end{equation}
In this expression, the first term is needed to cancel an
equivalent contribution coming from the Einstein-Hilbert action,
while the second term is responsible for the MOND force law that
one can read from (\ref{g00}).

We found that it was permissible, but not necessary, to involve a
second nonlocal scalar,
\begin{equation}
X[g] \equiv g^{\mu\nu} \partial_{\mu} \frac1{\Box} \Bigl[
u^{\alpha} u^{\beta} R_{\alpha\beta} \!-\! \frac12 R\Bigr]
\partial_{\nu} \frac1{\Box} \Bigl[u^{\rho} u^{\sigma}
R_{\rho\sigma} \!-\! \frac12 R\Bigr] \; . \label{Xdef}
\end{equation}
In the static, spherically symmetric and ultra-weak field limit
$X[g]$ reduces to $[b'(r) - a(r)/r]^2$. This means that a term
linear in $X[g]$ (with a suitable coefficient) in the action
would cancel, in the ultra-weak field limit, an analogous term in
the Einstein-Hilbert action responsible for (\ref{grr}) in the
$g_{rr}$ equation. To this term, one can then add a next-order
correction in the ultra-weak field expansion, such as
\begin{equation}
\Delta \mathcal{L}_x = \frac{a_0^2}{16 \pi G} \Biggl\{ -\frac12
\Bigl( \frac{X}{a_0^2} \Bigr) + \frac16 \Bigl( \frac{X}{a_0^2}
\Bigr)^{\frac32} + \dots \Biggr\} \sqrt{-g} \; . \label{DLX}
\end{equation}
Any successful implementation of MOND must involve the addition
of (\ref{DLY}), but the decision of whether or not to
additionally include (\ref{DLX}) is optional because
(\ref{grr}) holds both in general relativity and in the MOND
regime (see \cite{Deffayet:2011sk} for more details). Avoiding
deviations from existing tests of general relativity requires
that the higher order terms give suppression for large values of
$Y[g]/a_0^2$.

The purpose of this paper is to extend our past results by
deriving the MOND corrections to the field equations for a
general metric and then specialize them to the homogeneous and
isotropic geometry appropriate to cosmology. To keep the analysis
simple we define the scalar $\chi[g]$, whose normalized gradient
gives the timelike 4-velocity (\ref{udef}), using the same
inverse of the scalar d'Alembertian which appears in both $Y[g]$
and $X[g]$,
\begin{equation}
\chi[g] \equiv -\frac1{\Box} 1 \; . \label{chidef}
\end{equation}
We also consider the important changes which occur when one
alters the MOND acceleration $a_0$ from a fundamental constant to
a dynamical quantity which varies with the cosmological expansion
rate. As mentioned above, many authors have drawn attention to the
numerical coincidence $a_0 \approx c H_0/2\pi$ between the MOND
acceleration and the current value of the Hubble parameter
\cite{dynaccel}. With a timelike 4-velocity field such as
$u^{\mu}[g]$, whose divergence $D_\mu u^\mu = 3H$ in a
cosmological background, it is easy to make this relation dynamical
by the replacement
\begin{equation}
a_0 \longrightarrow \alpha[g] \equiv \frac{D_{\mu} u^{\mu}}{6\pi}
\; . \label{dyna0}
\end{equation}
In this way the extra MOND force, which is necessary if there is
no dark matter, can become effective even at early times during
which the condition $|Y/a_0^2| \gg 1$ would otherwise have
suppressed MOND effects.

This paper contains five sections of which this introduction is
the first. In section \ref{Y} we consider the simplest class of
models in which $\Delta \mathcal{L}$ depends only on the
invariant $Y[g]$, with the MOND acceleration $a_0$ a fundamental
constant. We derive the correction $\Delta G_{\mu\nu}$ to the
field equations for a general metric and then specialize this to
cosmology. Section \ref{X} carries out the same exercise for MOND
additions which also depend on the invariant $X[g]$, again with
constant $a_0$. In section \ref{a0} we derive the changes which
occur when the MOND acceleration is made dynamical through the
replacement (\ref{dyna0}). Section \ref{concl} gives our
conclusions.

\section{Models based on $Y$ with constant $a_0$}\label{Y}

The task of this section is to analyze the minimal class of
models,
\begin{equation}
\Delta \mathcal{L}_y = \frac1{16\pi G} \times a_0^2
f_y\Bigl(\frac{Y[g]}{a_0^2} \Bigr) \sqrt{-g} \; , \label{nonlocalY}
\end{equation}
where $Y[g]$ is the nonlocal invariant defined by expressions
(\ref{Ydef}), (\ref{udef}) and (\ref{chidef}), and $a_0$ is
strictly constant. We first express the nonlocal model
(\ref{nonlocalY}) in a local form involving the metric and four
auxiliary scalars. We next vary with respect to $g^{\mu\nu}$ to
derive the MOND addition to the Einstein tensor (\ref{DGmn}) for
a general metric, then specialize to the homogeneous, isotropic
and spatially flat geometry appropriate to cosmology. The section
closes with a discussion of how the function $f_y(Z)$ can be
chosen for $Z < 0$ (MOND phenomenology only fixes $f_y(Z)$ for $Z
> 0$) to support an arbitrary expansion history.

\subsection{General field equations}\label{genY}

We can derive causal and conserved field equations from the
nonlocal form (\ref{nonlocalY}) using the ``partial integration
trick'' of earlier studies \cite{SW,deser1,review}. However, it
is simpler to localize $\Delta \mathcal{L}_y$ using scalar
auxiliary fields after the procedure of Nojiri and Odintsov
\cite{Sergei}. Our model (\ref{nonlocalY}) requires scalars
$\phi$ and $\chi$ to stand for the two nonlocal expressions in
the original Lagrangian,
\begin{equation}
\phi \longrightarrow \frac2{\Box} u^{\alpha} u^{\beta}
R_{\alpha\beta} \qquad , \qquad \chi \longrightarrow -\frac1{\Box}
1 \; , \label{phichi}
\end{equation}
and Lagrange multiplier fields $\xi$ and $\psi$ to enforce these
relations. We shall abuse the notation slightly by employing the
same symbol for the local Lagrangian and its nonlocal ancestor
(\ref{nonlocalY}),
\begin{eqnarray}
\lefteqn{\Delta \mathcal{L}_y = \frac1{16\pi G} \Biggl\{ a_0^2 f_y
\Bigl(\frac{g^{\mu\nu} \partial_{\mu} \phi \partial_{\nu} \phi}{a_0^2}
\Bigr) } \nonumber \\
& & \hspace{2cm} - \Bigl[ \partial_{\mu} \xi \partial_{\nu} \phi
g^{\mu\nu} \!+\! 2 \xi R_{\mu\nu} u^{\mu} u^{\nu} \Bigr] - \Bigl[
\partial_{\mu} \psi \partial_{\nu} \chi g^{\mu\nu} \!-\! \psi\Bigr]
\Biggr\} \sqrt{-g} \; . \qquad \label{localY}
\end{eqnarray}
The 4-velocity field in this version of the model is still the
normalized gradient (\ref{udef}) of $\chi$, but the scalar $\chi$
is an independent variable.

It is straightforward to compute the MOND correction to the
Einstein tensor,
\begin{eqnarray}
\lefteqn{ \frac{16 \pi G}{\sqrt{-g}} \frac{\delta \Delta S_y}{\delta
g^{\mu\nu}} = \frac12 g_{\mu\nu} \Bigl[ -a_0^2 f_y + g^{\rho\sigma}
\Bigl( \partial_{\rho} \xi \partial_{\sigma} \phi \!+\!
\partial_{\rho} \psi \partial_{\sigma} \chi\Bigr) + 2 \xi
u^{\rho} u^{\sigma} R_{\rho\sigma} - \psi\Bigr] } \nonumber \\
& & \hspace{0cm} + \partial_{\mu} \phi \partial_{\nu} \phi f_y' -
\partial_{(\mu} \xi \partial_{\nu)} \phi - \partial_{(\mu} \psi
\partial_{\nu)} \chi - 2 \xi \Bigl[ 2 u_{(\mu} u^{\alpha}
R_{\nu ) \alpha}
\!+\! u_{\mu} u_{\nu} u^{\alpha} u^{\beta} R_{\alpha\beta} \Bigr]
\nonumber \\
& & \hspace{2.5cm} - \Bigl[ \Box (\xi u_{\mu} u_{\nu}) +
g_{\mu\nu} D_{\alpha} D_{\beta} (\xi u^{\alpha} u^{\beta}) -
2 D_{\alpha} D_{(\mu} (\xi u_{\nu)} u^{\alpha} ) \Bigr] \; . \qquad
\label{DGmnforY}
\end{eqnarray}
In this and subsequent expressions we follow the usual convention
in which parenthesized indices are symmetrized. Also, we denote
the covariant derivative with respect to $x^{\mu}$ by the symbol
$D_{\mu}$.

It remains to specify the various scalars as nonlocal functionals
of the metric. This follows applying retarded boundary conditions
to the field equations which result from varying (\ref{localY}),
\begin{eqnarray}
\frac{16 \pi G}{\sqrt{-g}} \, \frac{\delta \Delta S_y}{
\delta \xi} & = & \Box \phi - 2 u^{\alpha} u^{\beta} R_{\alpha\beta}
\; , \label{xieqn} \\
\frac{16 \pi G}{\sqrt{-g}} \, \frac{\delta \Delta S_y}{
\delta \psi} & = & \Box \chi + 1 \; , \label{psieqn} \\
\frac{16 \pi G}{\sqrt{-g}} \, \frac{\delta \Delta S_y}{
\delta \phi} & = & \Box \xi - 2 D_{\mu} \Biggl[ D^{\mu} \phi
f_y'\Bigl( \frac{g^{\rho\sigma} \partial_{\rho} \phi
\partial_{\sigma} \phi}{a_0^2} \Bigr) \Biggr] \; , \qquad
\label{phieqn} \\
\frac{16 \pi G}{\sqrt{-g}} \, \frac{\delta \Delta S_y}{
\delta \chi} & = & \Box \psi - 4 D_{\mu} \Biggl[ \frac{ \xi
g^{\mu\rho}_{\perp} u^{\sigma} R_{\rho\sigma}}{\sqrt{-g^{\alpha\beta}
\partial_{\alpha} \chi \partial_{\beta} \chi}} \Biggr] \; . \qquad
\label{chieqn}
\end{eqnarray}
(Note the induced metric $g^{\mu\nu}_{\perp} \equiv g^{\mu\nu} +
u^{\mu} u^{\nu}$ which appears in equation (\ref{chieqn}) for
$\psi$.) Solving for each of the four scalars involves inverting
scalar d'Alembertian $\Box$, which would ordinarily allow us to
freely specify each scalar and its first time derivative on the
initial value surface. Permitting those degrees of freedom would
result in two scalar ghosts \cite{deser2,review}. The original
nonlocal model is recovered by setting each scalar and its first
time derivative to zero on the initial value surface, which also
eliminates the ghosts (and in fact all the modes associated
with the scalars).

\subsection{Specialization to FLRW}\label{FLRWY}

On scales of $100~{\rm Mpc}$ and larger the geometry of our
universe is well described by a homogeneous, isotropic and
spatially flat metric in co-moving coordinates,
\begin{equation}
ds_{\scriptscriptstyle {\rm FLRW}}^2 \equiv g_{\mu\nu} dx^{\mu}
dx^{\nu} = -dt^2 + a^2(t) d\vec{x} \!\cdot\! d\vec{x} \; .
\label{FLRW}
\end{equation}
The function $a(t)$ is known as the scale factor and its
logarithmic time derivative gives the Hubble parameter $H(t)$,
\begin{equation}
H(t) \equiv \frac{\dot{a}}{a} \; . \label{Hubble}
\end{equation}
The nonvanishing components of the affine connection are,
\begin{equation}
\Gamma^i_{~j0} = H \delta^i_j \qquad , \qquad \Gamma^0_{~ij} =
H g_{ij} \; . \label{Gamma}
\end{equation}
This implies the following components of the curvature,
\begin{eqnarray}
R^0_{~i0j} = (\dot{H} + H^2) g_{ij} \quad , \quad R^i_{~jk\ell}
= H^2 (\delta^i_k g_{j\ell} - \delta^i_{\ell} g_{jk} ) \; , \\
R_{00} = -3 (\dot{H} + H^2) \quad , \quad R_{ij} = (\dot{H} +
3 H^2) g_{ij} \quad , \quad R = 6\dot{H}+ 12 H^2 \; .
\end{eqnarray}

The nonvanishing components of the second covariant derivative
of a scalar $S(t)$ are simple,
\begin{equation}
D_0 D_0 S = \ddot{S} \; , \; D_i D_j S = - H \dot{S} g_{ij}
\;\; \Longrightarrow \;\; \Box S = -(\ddot{S} + 3 H \dot{S})
= -\frac1{a^3} \frac{d}{dt} \, a^3 \dot{S} \; . \label{DDS}
\end{equation}
Of course the final expression for $\Box S$, with our retarded
boundary conditions, results in a simple form for
$\frac1{\Box} S$,
\begin{equation}
\Bigl[\frac1{\Box} \, S\Bigr](t) = -\int_{t_i}^t \!\!
\frac{dt'}{a^3(t')} \int_{t_i}^t \!\! dt'' a^3(t'') S(t'')
\; . \label{1overbox}
\end{equation}
We also require various contractions of double covariant
derivatives of a second rank tensor whose nonzero components are
restricted by homogeneity and isotropy to be $T_{00}(t)$ and
$T_{ij} = T(t) g_{ij}$. [Note that this $T$ does not mean the trace
of $T_{\mu\nu}$, and notably that it will vanish below for
$T_{\mu\nu} = \xi u_\mu u_\nu$.] Some tedious but straightforward
manipulations reveal,
\begin{eqnarray}
\Box T_{00} & = & -\ddot{T}_{00} - 3H \dot{T}_{00} + 6 H^2 (T_{00}
\!+\! T) \; , \\
\Box T_{ij} & = & \Bigl[-\ddot{T} - 3 H \dot{T} + 2 H^2 (T_{00}
\!+\! T)\Bigr] g_{ij} \; , \\
D_{\alpha} D_0 T^{\alpha}_{~0} & = & -\ddot{T}_{00} - 3 H
(\dot{T}_{00} \!+\! \dot{T}) + 3 H^2 (T_{00} \!+\! T) \; , \\
D_{\alpha} D_i T^{\alpha}_{~j} & = & \Bigl[ H \dot{T}_{00} +
(\dot{H} \!+\! 4 H^2) (T_{00} \!+\! T) \Bigr] g_{ij} \; , \\
D_{\alpha} D_{\beta} T^{\alpha\beta} & = & \ddot{T}_{00} + 3 H (2
\dot{T}_{00} \!+\! \dot{T}) + 3 (\dot{H} \!+\! 3 H^2) (T_{00} \!+\!
T) \; .
\end{eqnarray}

The various auxiliary fields take simple forms when specialized to
the FLRW geometry (\ref{FLRW}),
\begin{eqnarray}
\phi(t) & = & 6 \!\! \int_{t_i}^t \!\! \frac{dt'}{a^3(t')}
\int_{t_i}^{t'} \!\! dt'' a^3(t'') \Bigl[\dot{H}(t'') + H^2(t'')
\Bigr] \; , \label{FLRWphi} \\
\chi(t) & = & \!\! \int_{t_i}^t \!\! \frac{dt'}{a^3(t')}
\int_{t_i}^{t'} \!\! dt'' a^3(t'') \quad \Longrightarrow \quad
u^{\mu}(t) = \delta^{\mu}_0 \; , \qquad \label{FLRWchi} \\
\xi(t) & = & 2 \!\! \int_{t_i}^t \!\! dt' \dot{\phi}(t') f_y'\Bigl(-
\frac{\dot{\phi}^2(t')}{a_0^2} \Bigr) \; , \label{FLRWxi} \\
\psi(t) & = & 0 \; . \label{FLRWpsi}
\end{eqnarray}
Strictly speaking, $u^\mu$ is ill defined on the initial
value surface, because Eq.~(\ref{udef}) is singular when
$\dot\chi = 0$, but we can take the limit of the well-defined
$u^\mu(t)$ for $t\rightarrow t_i$. Alternative definitions
of this timelike unit vector may also be chosen, like
Eqs.~(20)--(22) of Ref.~\cite{Tsamis:2013cka}.

Of course homogeneity and isotropy imply that any second rank tensor
such as $\Delta G_{\mu\nu}$ has only two distinct components when
specialized to the FLRW geometry (\ref{FLRW}). We find them to be,
\begin{eqnarray}
\frac{16 \pi G}{\sqrt{-g}} \frac{\delta \Delta S_y}{\delta g^{00}}
\Biggl\vert_{\scriptscriptstyle {\rm FLRW}} \!\!
&=& \frac{a_0^2}2 f_y\Bigl( \frac{-\dot{\phi}^2}{a_0^2} \Bigr)
+ 3 H \dot{\xi} + 6 H^2 \xi \; , \label{00eqn} \\
\frac{16 \pi G}{\sqrt{-g}} \frac{\delta \Delta S_y}{\delta g^{ij}}
\Biggl\vert_{\scriptscriptstyle {\rm FLRW}} \!\!
& = &
-\Biggl[\frac{a_0^2}2 f_y\Bigl(\frac{-\dot{\phi}^2}{a_0^2}\Bigr) +
\ddot{\xi} + (\frac{\dot{\phi}}2 \!+\! 4 H) \dot{\xi} + (4\dot{H}
\!+\! 6 H^2) \xi\Biggr] g_{ij} \; . \qquad \label{ijeqn}
\end{eqnarray}

\subsection{The reconstruction problem}\label{reconY}

If the function $f_y(Z)$ in expression (\ref{nonlocalY}) were
known for $Z < 0$ then one would add expression (\ref{00eqn}) to
the usual Friedmann equation and solve for the scale factor
$a(t)$,
\begin{equation}
3 H^2 + \Biggl\{ \frac{a_0^2}2 f_y\Bigl( \frac{-\dot{\phi}^2}{a_0^2}
\Bigr) + 3 H \dot{\xi} + 6 H^2 \xi\Biggr\} = 8 \pi G \rho\,,
\label{friedman}
\end{equation}
where $\rho$ describes all matter sources, including radiation
and baryonic matter, but not dark matter nor dark energy which
would be reproduced by the nonlocal terms within the curly brackets.
However, MOND phenomenology only determines the asymptotic
forms of $f_y(Z)$ for $0 < Z \alt 1$ and for $Z \gg 1$,
\begin{eqnarray}
0 < Z \alt 1 & \Longrightarrow & f_y(Z) = \frac12 Z - \frac16
Z^{\frac32} + O(Z^2) \; , \\
1 \ll Z < \infty & \Longrightarrow & f_y(Z) \longrightarrow 0 \; .
\end{eqnarray}
The {\it reconstruction problem} consists of instead regarding
$a(t)$ as known --- along with how the energy density $\rho$
depends upon $a(t)$ --- and then solving the modified Friedmann
equation (\ref{friedman}) to find the function $f_y(Z)$ which
supports the desired expansion history (similarly to what has
been done in scalar-tensor theories \cite{reconTS} or other
nonlocal models \cite{recon}).

Once the reconstruction problem has been solved the model is
fixed, and one can subject it to meaningful tests by working
out its predictions for the growth of cosmological perturbations.
Many modified gravity models have been analyzed in this way. For
example, the free function $f(\frac1{\Box} R)$ of ``nonlocal
cosmology'' \cite{deser1,deser2} was determined (numerically) to
support the $\Lambda$CDM expansion history, without a
cosmological constant \cite{recon}, then its predictions for
structure formation were shown to be in conflict with the most
recent data on weak lensing and redshift space distortions
\cite{perts}. In this subsection we will derive a second order,
linear differential equation for $f_y(Z)$ which could be
numerically solved to support a given expansion history.

The first problem with (\ref{friedman}) is that time is the
natural variable, rather than $Z(t) \equiv
-[\dot{\phi}(t)/a_0]^2$. We therefore employ the new symbol
$f(t)$ to regard the dependent variable as a function of time,
\begin{equation}
f(t) \equiv f_y\Bigl(\frac{-\dot{\phi}^2(t)}{a_0^2} \Bigr) \qquad
\Longrightarrow \qquad f_y'\Bigl( \frac{-\dot{\phi}^2}{a_0^2}\Bigr)
= -\frac{a_0^2 \dot{f}}{2 \dot{\phi} \ddot{\phi}} \; .
\label{eq43}
\end{equation}
The second problem is that the auxiliary scalar $\xi$ given by
expression (\ref{FLRW}) involves an integral of $f$. We therefore
divide (\ref{friedman}) by $3 a_0^2 H^2(t)$ and differentiate,
\begin{equation}
\frac{d}{dt} \Bigl[ -\frac{\dot{f}}{H \ddot{\phi}}
+ \frac{f}{6 H^2}\Bigr]
- \frac{2\dot{f}}{\ddot{\phi}} = \frac{d}{dt}
\Bigl[ \frac{8\pi G \rho}{3
a_0^2 H^2} \Bigr] \; . \label{2ndorder}
\end{equation}
Equation (\ref{2ndorder}) is a linear, second order differential
equation for $f(t)$ which can be evolved forward from $t = t_i$,
using the explicit expression (\ref{FLRWphi}) for $\phi(t)$ in
terms of the known expansion history. From the mathematical point
of view, $f(t)$ is fully determined from the two initial
conditions $\dot f(t_i) = 0$ and $f(t_i) = 2\left[8\pi G
\rho(t_i)-3 H(t_i)^2\right]/a_0^2$, implied by
Eqs.~(\ref{FLRWphi}), (\ref{FLRWxi}), (\ref{friedman}) and
(\ref{eq43}). However, we actually have more freedom because
$H(t)$ is not known with infinite precision. At early times
during radiation domination, we only need our nonlocal terms,
within the curly brackets of Eq.~(\ref{friedman}), to be
negligible with respect to $8\pi G \rho_{\text{radiation}}$. As
will be detailed in a forthcoming publication, it is thus
possible to integrate (\ref{2ndorder}) backwards in time,
starting from the present epoch, while still integrating forward
(\ref{FLRWphi}) and (\ref{FLRWxi}) to respect the crucial
constraints $\phi(t_i) = \dot\phi(t_i) = \xi(t_i) = \dot\xi(t_i)
= 0$ which eliminate ghost excitations. One of the two
integration constants in the solution of (\ref{2ndorder}) is then
fixed by requiring that the undifferentiated equation
(\ref{friedman}) holds, while the second constant allows us to
match the limit of $f_y(Z)$ for $Z\rightarrow 0^-$ to the needed
MOND form (\ref{DLY}) for $Z > 0$.

The final step is inverting the relation between $Z$ and
$t$ implied by relation (\ref{FLRWphi}) to solve for $t$ as a
function of $Z$,
\begin{equation}
Z(t) = - \Biggl[ \frac6{a_0\,a^3(t)} \int_{t_i}^t \!\! dt'
a^3(t') \Bigl[
\dot{H}(t') + H^2(t')\Bigr] \Biggr]^2 \qquad \Longrightarrow \qquad
t(Z) \; .
\end{equation}
The desired function is $f_y(Z) = f(t(Z))$. Except for very
simple expansion histories this analysis will need to be done
numerically.

\section{Models which include $X$ with constant $a_0$}\label{X}

The purpose of this section is to work out how $\Delta
G_{\mu\nu}$ changes if, in addition to the mandatory MOND term
(\ref{nonlocalY}) we elect to also add the optional term,
\begin{equation}
\Delta \mathcal{L}_x = \frac1{16\pi G} \times a_0^2
f_x\Bigl(\frac{X[g]}{a_0^2} \Bigr) \sqrt{-g} \; , \label{nonlocalX}
\end{equation}
where $X[g]$ is the nonlocal invariant defined by expressions
(\ref{Xdef}), (\ref{udef}) and (\ref{chidef}), and $a_0$ is
strictly constant. Much of the analysis is similar to what was
done in section \ref{Y} for $\Delta \mathcal{L}_y$. In
particular, it is again useful to localize the system using
auxiliary scalar fields. In addition to $\chi$ and $\psi$ ---
which are already present in the mandatory term (\ref{localY})
--- we require a scalar $\theta$ which bears the same relation to
$X[g]$ that $\phi$ bears to $Y[g]$,
\begin{equation}
\theta \longrightarrow \frac1{\Box} \Bigl[ u^{\alpha} u^{\beta}
R_{\alpha\beta} - \frac12 R\Bigr] \; .
\end{equation}
Of course we also need a Lagrange multiplier field $\omega$ to
enforce this relation. This makes the local version,
\begin{equation}
\Delta \mathcal{L}_x = \frac1{16\pi G} \Biggl\{ a_0^2 f_x\Bigl(
\frac{g^{\mu\nu} \partial_{\mu} \theta \partial_{\nu} \theta}{a_0^2}
\Bigr) - \partial_{\mu} \omega \partial_{\nu} \theta g^{\mu\nu} -
\omega \Bigl[u^{\alpha} u^{\beta} R_{\alpha\beta} \!-\! \frac12 R
\Bigr] \Biggr\} \sqrt{-g} \; . \label{localX}
\end{equation}
We remind the reader that $u^{\mu}$ is the normalized gradient
(\ref{udef}) of $\chi$, which appears, along with its Lagrange
multiplier $\psi$, in the mandatory term (\ref{localY}).

The correction (\ref{localX}) makes to $\Delta G_{\mu\nu}$ is,
\begin{eqnarray}
\lefteqn{ \frac{16 \pi G}{\sqrt{-g}} \frac{\delta \Delta S_x}{
\delta g^{\mu\nu}} = \frac12 g_{\mu\nu} \Bigl[- a_0^2 f_x \!+\!
\partial_{\rho} \omega \partial_{\sigma} \theta g^{\rho\sigma}
\!+\! \omega \Bigl(u^{\alpha} u^{\beta} R_{\alpha\beta} \!-\!
\frac12 R\Bigr) \Bigr] \!+\! \partial_{\mu} \theta \partial_{\nu}
\theta f_x' } \nonumber \\
& & \hspace{0cm} - \partial_{(\mu} \omega \partial_{\nu)} \theta
- \omega \Bigl[2 u_{(\mu} u^{\alpha} R_{\nu) \alpha} \!+\! u_{\mu}
u_{\nu} u^{\alpha} u^{\beta} R_{\alpha\beta} \!-\! \frac12
R_{\mu\nu} \Bigr] + \frac12 (g_{\mu\nu} \Box \!-\! D_{\mu}
D_{\nu}) \omega \nonumber \\
& & \hspace{2cm} - \frac12 \Bigl[\Box (\omega u_{\mu}
u_{\nu}) \!+\! g_{\mu\nu} D_{\alpha} D_{\beta} (\omega u^{\alpha}
u^{\beta}) \!-\! 2 D_{\alpha} D_{(\mu} (\omega u_{\nu)}
u^{\alpha})\Bigr] \; . \qquad \label{DGmnX}
\end{eqnarray}
The auxiliary fields $\theta$ and $\omega$ are determined by
applying retarded boundary conditions to the equations which
derive from varying $\Delta S_x$,
\begin{eqnarray}
\frac{16 \pi G}{\sqrt{-g}} \, \frac{\delta \Delta S_x}{
\delta \omega} & = & \Box \theta - \Bigl[u^{\alpha} u^{\beta}
R_{\alpha\beta} -\frac12 R\Bigr] \; , \label{omegaeqn} \\
\frac{16 \pi G}{\sqrt{-g}} \, \frac{\delta \Delta S_x}{
\delta \theta} & = & \Box \omega - 2 D_{\mu} \Biggl[ D^{\mu} \theta
f_x'\Bigl( \frac{g^{\rho\sigma} \partial_{\rho} \theta
\partial_{\sigma} \theta}{a_0^2} \Bigr) \Biggr] \; . \qquad
\label{thetaeqn}
\end{eqnarray}
The equation for the auxiliary field $\chi$ is still
(\ref{psieqn}), but the equation for $\psi$ receives
contributions from the $\chi$ dependence (through $u^{\mu}$) in
both $\Delta \mathcal{L}_y$ and $\Delta \mathcal{L}_x$,
\begin{equation}
\frac{16 \pi G}{\sqrt{-g}} \, \frac{\delta \Delta S_{x+y}}{
\delta \chi} = \Box \psi - 2 D_{\mu} \Biggl[ \frac{ (2\xi \!+\!
\omega) g^{\mu\rho}_{\perp} u^{\sigma} R_{\rho\sigma}}{
\sqrt{-g^{\alpha\beta} \partial_{\alpha} \chi \partial_{\beta} \chi}}
\Biggr] \; . \qquad \label{newchieqn}
\end{equation}
Just as for the mandatory addition $\Delta \mathcal{L}_y$, we do
not regard the auxiliary scalars as fundamental fields with
arbitrary initial value data. That would result in the
combination $\theta - \omega$ being a ghost \cite{deser2,review}.
We instead define each auxiliary field and its first time
derivative to vanish at $t = t_i$.

Specializing to the FLRW geometry is a straightforward extension
of the analysis of subsection \ref{FLRWY}. The addition of the
optional term (\ref{nonlocalX}) does not change the FLRW
expressions (\ref{FLRWphi}-\ref{FLRWpsi}) for the four auxiliary
scalars $\phi$, $\chi$, $\xi$ and $\psi$ of the mandatory MOND
term (\ref{nonlocalY}). [Note in particular that $\psi$ still
vanishes identically, Eq.~(\ref{FLRWpsi}), because
$g^{\mu\rho}_{\perp} u^{\sigma} R_{\rho\sigma} =
g^{\mu\rho}_{\perp} R_{\rho 0} =
g^{\mu 0}_{\perp} R_{00} = 0$ in FLRW.]
The two new auxiliary scalars become,
\begin{eqnarray}
\theta(t) & = & \int_{t_i}^t \!\! \frac{dt'}{a^3(t')}
\int_{t_i}^{t'} \!\! dt'' a^3(t'') \Bigl[6\dot{H}(t'') + 9 H^2(t'')
\Bigr] \; , \label{FLRWtheta} \\
\omega(t) & = & 2 \!\! \int_{t_i}^t \!\! dt' \dot{\theta}(t')
f_x'\Bigl(-
\frac{\dot{\theta}^2(t')}{a_0^2} \Bigr) \; . \label{FLRWomega}
\end{eqnarray}
And the contributions to the two nonzero components of $\Delta
G_{\mu\nu}$ are,
\begin{eqnarray}
\frac{16 \pi G}{\sqrt{-g}} \frac{\delta \Delta S_x}{\delta g^{00}}
\Biggl\vert_{\scriptscriptstyle {\rm FLRW}} \!\!
& = & \frac{a_0^2}2 f_x\Bigl( \frac{-\dot{\theta}^2}{a_0^2}
\Bigr)
+ 3 H \dot{\omega} + \frac92 H^2 \omega \; , \label{X00eqn} \\
\frac{16 \pi G}{\sqrt{-g}} \frac{\delta \Delta S_x}{\delta g^{ij}}
\Biggl\vert_{\scriptscriptstyle {\rm FLRW}} \!\!
& = & -\Biggl[\frac{a_0^2}2 f_x
\Bigl(\frac{-\dot{\theta}^2}{a_0^2}
\Bigr) + \ddot{\omega} + \left(\frac{\dot{\theta}}2 \!+\! 3 H\right)
\dot{\omega} + \left(3\dot{H} \!+\! \frac92 H^2\right) \omega\Biggr]
g_{ij} \; . \qquad \label{Xijeqn}
\end{eqnarray}

An important observation is that $X$ vanishes for exact matter
domination, $H(t) = \frac2{3t}$. This means that the optional
correction cannot have much effect on cosmology at the time of
recombination, or on the early stages of structure formation. It
also means that the optional correction cannot supply the MOND
enhancement of gravity which would be necessary to compensate for
the absence of dark matter at early times.

\section{Making $a_0$ dynamical}\label{a0}

MOND phenomenology only constrains the function $f_y(Z)$ of the
mandatory MOND addition (\ref{nonlocalY}) for $Z > 0$. Based on
subsection \ref{reconY}, it seems possible to adjust how $f_y(Z)$
behaves for $Z < 0$ to support an arbitrary expansion history.
However, the variable $Y[g]$ is essentially $-H^2(t)$ for
cosmology, whereas $a_0 \sim H_0/2\pi$, so the argument $Z =
Y/a_0^2 \sim -4\pi^2 H^2/H_0^2$ varies enormously over
interesting cosmological events such as nucleosynthesis ($Z \sim
-10^{32}$) and recombination ($Z \sim - 10^{10}$). This raises
concerns about fine tuning. These concerns can be ameliorated by
making the MOND acceleration $a_0$ some functional $\alpha[g]$ of
the metric so that it changes with the scale of cosmological
acceleration. There are many, many plausible choices for
$\alpha[g]$. To develop some quantitative understanding of the
consequences of a sliding scale, we here work out the effect of a
simple choice (\ref{dyna0}) in which $a_0$ is replaced by
$1/6\pi$ times the expansion, i.e., $D_\mu u^\mu/6\pi$ where the
timelike 4-velocity $u^{\mu}[g]$ is defined in Eqs.~(\ref{udef})
and (\ref{chidef}). Because the optional MOND addition
(\ref{nonlocalX}) does not seem to have much effect for cosmology
we only derive results for the mandatory addition
(\ref{nonlocalY}).

The replacement (\ref{dyna0}) causes only three changes in the
general metric field equations (\ref{DGmnforY}) and
(\ref{xieqn}-\ref{chieqn}). The first and most obvious change is
that the factors of $a_0$ in (\ref{DGmnforY}) and (\ref{phieqn})
get replaced by $\alpha[g]$. The second change is that the
addition to the Einstein tensor acquires an extra contribution
from the metric dependence of $\alpha[g]$,
\begin{eqnarray}
\Bigl(\Delta G_{\mu\nu} \Bigr)_{\rm new} &=& \Bigl( \Delta G_{\mu\nu}
\Bigr)_{\rm old} + g_{\mu\nu} \Bigl[\alpha^2 f_y - g^{\rho\sigma}
\partial_{\rho} \phi \partial_{\sigma} \phi f_y'\Bigr]\nonumber\\
&&+\frac1{6\pi} \Bigl[g_{\mu\nu} u^\gamma \partial_\gamma
-2 u_{(\mu} \partial_{\nu)} - u_{\mu}
u_{\nu} u^{\gamma} \partial_{\gamma}\Bigr] \Bigl[ \alpha f_y
- \frac1{\alpha}
g^{\rho\sigma} \partial_{\rho} \phi \partial_{\sigma} \phi
f_y'\Bigr] \; .
\qquad \label{newGmn}
\end{eqnarray}
The final change is that equation (\ref{chieqn}) for the
auxiliary scalar $\psi$ picks up an extra term from the $\chi$
dependence of $\alpha$,
\begin{equation}
\frac{16 \pi G}{\sqrt{-g}} \, \frac{\delta \Delta S_y}{\delta \chi}
= \Box \psi - D_{\mu} \Biggl[ \frac{4 \xi g^{\mu\nu}_{\perp} u^{\rho}
R_{\rho\nu} \!+\!
\frac1{3\pi} g^{\mu\nu}_{\perp} \partial_{\nu}
[\alpha f_y \!-\! \frac1{\alpha}
g^{\rho\sigma} \partial_{\rho} \phi \partial_{\sigma} \phi f_y']}{
\sqrt{-g^{\kappa\lambda} \partial_{\kappa} \chi
\partial_{\lambda} \chi}} \Biggr]\; . \label{newchi}
\end{equation}

These small alterations in the functional form of the field
equations conceal vast changes in their numerical values. That
becomes apparent upon specialization to the FLRW cosmology
(\ref{FLRW}). In this case the functional $\alpha[g]$ becomes,
\begin{equation}
\alpha[g] \Biggl\vert_{\scriptscriptstyle {\rm FLRW}}
= \frac{H(t)}{2\pi} \; .
\end{equation}
The auxiliary scalars $\phi$, $\chi$ and $\psi$ are unchanged
from expressions (\ref{FLRWphi}), (\ref{FLRWchi}) and
(\ref{FLRWpsi}), respectively, but $\xi$ becomes,
\begin{equation}
\xi(t) = 2 \int_{t_i}^t \!\! dt' \dot{\phi}(t') f_y'\Bigl(
\frac{-4\pi^2 \dot{\phi}^2(t')}{H^2(t')} \Bigr) \; . \label{newxi}
\end{equation}
Our nonlocal addition to the Friedmann equation is
\begin{equation}
\Delta G_{00} = -\frac{H^2}{8\pi^2} f_y - \dot{\phi}^2 f_y' +
3 H \dot{\xi} + 6 H^2 \xi \; , \label{newFried}
\end{equation}
where each of the functions $f_y(Z)$ is evaluated at $Z = -4\pi^2
\dot{\phi}^2/H^2$.

Let us underline two subtleties related to the way we introduce
a time-dependent $a_0$ through the replacement
$a_0 \rightarrow \alpha[g]$. First of all, in a FLRW background,
the argument $Z$ of the function $f_y(Z)$ will remain of the
order of $-4\pi^2$ at all times, so that the reconstruction of an
arbitrary expansion history seems more difficult to achieve.
This needs to be analyzed numerically. There are however
many other possible definitions of a time-dependent $a_0$,
and significant but not-too-large variations of Z are possible
for instance with some kind of geometrical mean between
$\alpha[g]$ and the constant $a_0$. Independently of the
cosmological reconstruction, corresponding to $Z<0$, note
that a time-dependent $a_0 \rightarrow \alpha[g]$ will be
quite useful in the MOND correction (\ref{DLY}) for $Z > 0$,
so that this modified dynamics happen for larger accelerations
at earlier times, mimicking the clustering effects of dark matter.

The second subtlety is that our definition (\ref{dyna0}) for
$\alpha[g]$ is likely to vanish within gravitationally bound
systems, because the local expansion $D_\mu u^\mu$ should
not keep any information about the asymptotic cosmological
evolution (although this needs to be confirmed by further
examination). This would force the model always into the
general relativistic regime, which would turn off the MOND
force even in the static, spherically symmetric and ultra-weak
field regime! One might deal with this by simply using
$a_0 + \alpha[g]$ as the acceleration scale \cite{Milgrom:2008cs}
entering our nonlocal action, or one might devise a more
nonlocal version of $\alpha[g]$ whose value inside a
gravitationally bound system depends upon the
cosmological expansion around it~\cite{Kiselev:2011fs}.

\section{Conclusions}\label{concl}

Our previous work on extending MOND to a relativistic, metric
theory led to consideration of two nonlocal scalar functionals of
the metric: a mandatory one $Y[g]$ given in expression
(\ref{Ydef}) and and optional one $X[g]$ given in expression
(\ref{Xdef}) \cite{Deffayet:2011sk}. To recover MOND with
sufficient weak lensing requires that the $Y$ term, and allows
that the $X$ term, be added to the gravitational Lagrangian in
the form,
\begin{equation}
\frac{\Delta \mathcal{L}}{\sqrt{-g}} = \frac{a_0^2}{16\pi G}
\Bigl[ f_y\Bigl( \frac{Y[g]}{a_0^2} \Bigr)
+ f_x\Bigl( \frac{X[g]}{a_0^2}\Bigr) \Bigr]
= \frac1{32\pi G} \Bigl[(Y - X) - \frac{(Y^{\frac32}
- X^{\frac32})}{3 a_0} + \dots \Bigr] \; .
\end{equation}
Our previous study gave the field equations for static,
spherically symmetric geometries in the ultra-weak field limit.
In this paper we have derived the field equations --- expressed
as an addition $\Delta G_{\mu\nu}$ to the usual Einstein tensor
--- for arbitrary functions $f_y$ and $f_x$, and for an arbitrary
metric. Our result for $\Delta G_{\mu\nu}$ from the mandatory
term is equation (\ref{DGmnforY}), with auxiliary fields
(\ref{xieqn}-\ref{chieqn}). Our result for $\Delta G_{\mu\nu}$
from the optional term is equation (\ref{DGmnX}), with auxiliary
fields (\ref{omegaeqn}-\ref{newchieqn}).

We also specialized the general field equations to the FLRW
geometry (\ref{FLRW}) of cosmology. Our results for the mandatory
term are relations (\ref{FLRWphi}-\ref{ijeqn}); for the optional
term they are relations (\ref{FLRWtheta}-\ref{Xijeqn}). Because
$X$ happens to vanish for a matter-dominated cosmology, the
optional term does not seem likely to play much role in
cosmology. However, in subsection \ref{reconY} we described a
technique by which the free function $f_y(Z)$ could be
constructed for $Z < 0$ --- which is not constrained by MOND
phenomenology --- to support a general expansion history $a(t)$.

Although the reconstruction problem can be solved for the
mandatory term, there will be large variations in the argument $Z
= -\dot{\phi}^2/a_0^2 \sim -4\pi^2 H^2/H_0^2$ over the course of
cosmological history. This makes it likely that the extra MOND
force --- which is needed at early times if there is no dark
matter --- will only become effective at recent times. That
argues for making $a_0$ dynamical. In section \ref{a0} we derived
the field equations and their specialization to cosmology for a
simple ansatz (\ref{dyna0}) in which the MOND constant $a_0$
changes with the expansion of the universe. One obvious problem
is that our definition (\ref{dyna0}) for the dynamical MOND
acceleration $\alpha[g]$ probably vanishes inside a
gravitationally bound structure, so that the MOND force
would always be turned off. We conclude that a more nonlocal
ansatz may be necessary, in which the MOND acceleration inside
gravitationally bound structures can still be determined by the
cosmological expansion around them.

\section*{Acknowledgments}
We have benefited from conversation and correspondence on this
subject with S. Deser, C. Skordis and M. Milgrom. This work was
partially supported by the European Research Council under the
European Community's Seventh Framework Programme (FP7/2007-2013
Grant Agreement no. 307934, NIRG project), by the ANR THALES
grant, by NSF grant PHY-1205591, and by the Institute for
Fundamental Theory at the University of Florida.

\end{document}